\documentclass[fdp,a4paper,fleqn%
]{w-art}
\usepackage{times,cite,w-thm}
\theoremstyle{plain}

\theoremstyle{definition}

\usepackage[]{graphicx}
\usepackage{color}
\def\beq{\begin{equation}}
\def\eeq{\end{equation}}
\def\beqar{\begin{eqnarray}}
\def\eeqar{\end{eqnarray}}
\def\drbar{\ifmmode{\overline{\rm DR}} \else{$\overline{\rm DR}$} \fi}

\newcommand{\be}[1]{\begin{equation} \label{(#1)}}
\newcommand{\ee}{\end{equation}}
\newcommand{\baq}[1]{\begin{eqnarray} \label{(#1)}}
\newcommand{\eaq}{\end{eqnarray}}

\def\lsim{\raise0.3ex\hbox{$\;<$\kern-0.75em\raise-1.1ex\hbox{$\sim\;$}}}
\def\gsim{\raise0.3ex\hbox{$\;>$\kern-0.75em\raise-1.1ex\hbox{$\sim\;$}}}

\def\beq   {\begin{equation}}
\def\eeq   {\end{equation}}
\def\beqd  {\begin{displaymath}}
\def\eeqd  {\end{displaymath}}
\def\beqaa {\begin{eqnarray}}
\def\eeqaa {\end{eqnarray}}

\def\ti  {\tilde}

\def\sf  {\ti f}

\def\su  {\ti u}

\def\sc  {\ti c}

\def\st  {\ti t}

\def\sg  {\ti g}

\def\sl  {\ti \ell}

\def\nt  {\tilde\chi^0}
\def\ch  {\tilde\chi^\pm}

\def\a   {\alpha}
\def\b   {\beta}

\def\gev {\rm ~GeV}

\def\sz{\ifmmode{\tilde{\chi}^0} \else{$\tilde{\chi}^0$} \fi}
\def\sw{\ifmmode{\tilde{\chi}} \else{$\tilde{\chi}$} \fi}

\begin{document}
\DOIsuffix{theDOIsuffix}
\Volume{55}
\Month{01}
\Year{2007}
\pagespan{1}{}

\keywords{Supersymmetry, MSSM,  squarks, flavour violation, LHC}
\subjclass[pacs]{11.30.Pb, 11.30.Hv, 12.60.Jv, 13.85.-t,  14.80.Ly
\qquad\parbox[t][2.2\baselineskip][t]{100mm}{%
\raggedright
\vfill}}%
  



\title[Squark generation mixing at LHC]{Flavour violating up-squark decays at LHC}


%
\author[H. Eberl]{Helmut Eberl\inst{1}%
\footnote{Speaker, e-mail:~\textsf{helmut@hephy.oeaw.ac.at}, Phone:
    +43\,(0)1\,544732835\,,  Fax: +43\,(0)1\,544732854}}
\address[\inst{1}]{Institut f\"ur Hochenergiephysik der \"OAW, Nikolsdorfer Gasse 18, 1050 Wien, Austria}
\author[A. Bartl]{A. Bartl\inst{2}}
\address[\inst{2}]{Fakult\"at f\"ur Physik, Universit\"at Wien,  A-1090 Vienna, Austria}
\author[B. Herrmann]{B. Herrmann\inst{3}}
\address[\inst{3}]{Deutsches Elektronen-Synchrotron (DESY), Theory Group, 
D-22603 Hamburg, Germany}
\author[K. Hidaka]{K. Hidaka\inst{4}}
\address[\inst{4}]{Department of Physics, Tokyo Gakugei University, Koganei, 
Tokyo 184-8501, Japan}
\author[W. Majerotto]{W. Majerotto\inst{1}}
\author[W. Porod]{W. Porod\inst{5}}
\address[\inst{5}]{Institut f\"ur Theoretische Physik und Astrophysik, Universit\"at W\"urzburg, 
D-97074 W\"urzburg, Germany}


 \begin{abstract}
We study the effect of squark generation mixing on squark production and decays at the LHC
in the Minimal Supersymmetric Standard Model (MSSM).
We show that the effect can be very large despite the very strong constraints
on quark flavour violation (QFV) from experimental data on B mesons.
We find that the two lightest up-type squarks ${\tilde u}_{1,2}$ can have large branching
ratios for the decays into $c  {\tilde\chi_1^0}$ and $t  {\tilde\chi_1^0}$ at the same
time, leading to QFV signals 
'$p p \to c {\bar t}\, (t {\bar c})$ + missing-$E_T$ + $X$' with a significant rate.
The observation of this remarkable signature would provide a powerful test of 
supersymmetric QFV at LHC. This could have a significant impact on the search for squarks
and the determination of the underlying MSSM parameters.
\end{abstract}
\maketitle                   
 
\vspace*{-10mm}
\section{Introduction}

The Minimal Supersymmetric Standard Model is the most intensively studied extension 
of the Standard Model (SM) of elementary particles. Low energy Supersymmetry (SUSY)
provides a solution to the so-called hierarchy problem in the Higgs sector, allows gauge
coupling unification, and provides possible candidates for the dark matter observed in our
Universe. If the MSSM is realized in nature, LHC will copiously produce gluinos and squarks, 
the supersymmetric partners of gluons and quarks, for masses of $O(1~TeV)$.
However, even if SUSY is discovered, it will be still a long way to determine the parameters
of the underlying model, which would shed light on the mechanism of SUSY breaking. 
So-called soft-SUSY-breaking terms are introduced in the SUSY Lagrangian.
As these soft-SUSY-breaking terms could also be the source of flavour violation beyond 
the Standard Model (SM), the measurement of flavour violating observables is directly linked to the 
crucial question about the SUSY-breaking mechanism. 
It is usually assumed that production and decays of squarks are 
quark-flavour conserving (QFC). However, additional flavour structures 
(i.e.~squark generation mixings) would imply that squarks are not quark-flavour 
eigenstates, which could result in sizable quark-flavour violation (QFV) effects 
significantly larger than those due to the Cabibbo-Kobayashi-Maskawa (CKM) mixing. 
In this contribution we study the effect of QFV due to the mixing of charm-squarks and top-squarks on production
and decays of squarks at the LHC in the general MSSM.

\section{Squark mixing with flavour violation and constraints}

The most general up-type squark mass matrix including left-right mixing
as well as quark-flavour mixing in the super-CKM basis of 
$\tilde u_{0\gamma}=
(\tilde u_L,\tilde c_L,\tilde t_L,\tilde u_R,\tilde c_R,\tilde t_R)$, 
$\gamma=1,\dots,6$, is \cite{Bartl:QFV_squark}

\baq{eq:SquarkMassMatrix}
M^2_{\tilde u}=\left(\begin{array}{ccc}
M^2_{\tilde u LL} & (M^2_{\tilde u RL})^\dagger\\[5mm]
M^2_{\tilde u RL} & M^2_{\tilde u RR}
\end{array}\right)~,\quad {\rm where~the~three~}  3\times3 {\rm ~matrices~read}
\eaq
\begin{eqnarray}
(M^2_{\tilde u LL})_{\alpha\beta} & = & 
M^2_{Q_u \alpha\beta}+\left[(\frac{1}{2}-\frac{2}{3}\sin^2\theta_W)
\cos2\beta~m_Z^2+m_{u_\alpha}^2\right]\delta_{\alpha\beta},
\label{eq:LL}\\[3mm]
(M^2_{\tilde u RR})_{\alpha\beta} & = & M_{U\alpha\beta}^2
+\left[\frac{2}{3}\sin^2\theta_W\cos2\beta~
m_Z^2+m_{u_\alpha}^2\right] \delta_{\alpha\beta}~,
\label{eq:RR}\\[3mm]
(M^2_{\tilde u RL})_{\alpha\beta} & = & (v_2/\sqrt{2} ) T_{U\beta\alpha}-
m_{u_\alpha} \mu^*\cot\beta~\delta_{\alpha\beta}~.
\label{eq:RL}
\end{eqnarray}
The indices $\alpha,\beta=1,2,3$ characterize the quark flavours $u,c,t$, respectively.
$M_{Q_u}^2$ and $M_U^2$ are the hermitean soft-SUSY-breaking mass matrices for the left 
and right up-type squarks, respectively. 
$T_U$ is the soft-SUSY-breaking trilinear coupling matrix of the up-type squarks.
The physical mass eigenstates $\tilde u_i$, $i=1,\dots,6$, are given
by $\tilde u_i=R^{\tilde u}_{i\alpha}\tilde u_{0\alpha}$. 
The mixing matrix $R^{\tilde u}$ and the mass eigenvalues 
are obtained by a unitary transformation 
$R^{\tilde u} M^2_{\tilde u} R^{\tilde u\dagger}=
{\rm diag}(m^2_{\tilde u_1},\dots,m^2_{\tilde u_6})$, where $m_{\tilde u_i}<m_{\tilde u_j}$
for $i<j$. \\
We define dimensionless QFV parameters 
$\delta^{uLL}_{\alpha\beta}$ and $\delta^{uRR}_{\alpha\beta}$ as  
\begin{eqnarray}
\delta^{uLL}_{\alpha\beta}  \equiv  M^2_{Q \alpha\beta} / \sqrt{M^2_{Q \alpha\alpha} M^2_{Q \beta\beta}}~,
\label{eq:InsLL}\qquad
\delta^{uRR}_{\alpha\beta} \equiv M^2_{U \alpha\beta} /  \sqrt{M^2_{U \alpha\alpha} M^2_{U \beta\beta}}~,
\label{eq:InsRR}
\end{eqnarray}
for $\delta^{uRL}_{\alpha\beta}$ see \cite{Bartl:QFV_squark}.
For example, $\delta^{uLL}_{23}$ and $\delta^{uRR}_{23}$ are the $\sc_L - \st_L$ and $\sc_R-\st_R$
mixing parameters.
In our analysis, we impose the following conditions
on the MSSM parameter space in order to respect experimental
and theoretical constraints, for details see \cite{Bartl:QFV_squark}:
{\renewcommand{\labelenumi}{\roman{enumi})}
\begin{enumerate}
\item 
Constraints from the B-physics experiments\cite{Bartl:refstherein}, such as
$2.92 \times 10^{-4} < B(b \to s ~\gamma) < 4.22 \times  10^{-4}$, 
$|\Delta M_{B_s}^{\rm SUSY} - 17.77|  <  3.31 ~{\rm ps}^{-1}$ and those from other
rare B-decays

\item
The experimental limit on the SUSY contributions to the $\rho$ parameter

\item
LEP and Tevatron lower limits on the SUSY particle masses

\item 
Vacuum stability conditions for the elements of the trilinear coupling matrices $T_U$ and $T_D$
\end{enumerate}
}

The conditions i) and iv) strongly constrain the 2$^{\rm nd}$ and 3$^{\rm rd}$ generation squark mixing
parameters $M^2_{Q23}$, $M^2_{D23}$, $T_{U23}$, $T_{D23}$, and $T_{D32}$, but not
$M^2_{U23}$ which is parametrized by $\delta^{uRR}_{23}$.
In the following numerical presentation all of these constraints are fulfilled. 

\section{Flavour violating fermionic squark decays at LHC}

First we study the effect of the mixing between the 2$^{\rm nd}$ and 3$^{\rm rd}$ generation of up-squarks on their decays only. The widths of the two--body decays
\baq{eq:decaymodes}
\tilde u_i &\to& u_k ~\sg, ~u_k ~\tilde\chi^0_n, ~d_k ~\tilde\chi^+_m, ~
\tilde{u}_j~Z^0, ~\tilde{d}_j~W^+, ~\tilde{u}_j~h^0
\eaq
are calculated, 
where $u_k=(u,c,t)$ and $d_k=(d,s,b)$.
A specific numerical scenario is chosen so that QFV signals at LHC are large.
It is given in Table \ref{eberl:tab1}.
\begin{table}[t]
\begin{center}
\begin{tabular}{|c||c|c|c|} \hline
 $M^2_{Q\alpha\beta}$
& \multicolumn{1}{c|}{\scriptsize{${\beta=1}$}} 
& \multicolumn{1}{c|}{\scriptsize{$\beta=2$}} 
& \multicolumn{1}{c|}{\scriptsize{$\beta=3$}} \\\hline\hline
 \scriptsize{$\alpha=1$}
& \multicolumn{1}{c|}{$(920)^2$} 
& \multicolumn{1}{c|}{0} 
& \multicolumn{1}{c|}{0} \\\hline

 \scriptsize{$\alpha=2$}
& \multicolumn{1}{c|}{0} 
& \multicolumn{1}{c|}{$(880)^2$} 
& \multicolumn{1}{c|}{$(224)^2$} \\\hline

 \scriptsize{$\alpha=3$}
& \multicolumn{1}{c|}{0} 
& \multicolumn{1}{c|}{$(224)^2$} 
& \multicolumn{1}{c|}{$(840)^2$} \\\hline
\end{tabular}
\begin{tabular}{|c|c|c|c|c|c|} \hline 
  \multicolumn{1}{|c|}{$M_1$} 
& \multicolumn{1}{c|}{$M_2$} 
& \multicolumn{1}{c|}{$M_3$} 
& \multicolumn{1}{c|}{$\mu$} 
& \multicolumn{1}{c|}{$\tan\beta$} 
& \multicolumn{1}{c|}{$m_{A^0}$} \\\hline\hline
 
  \multicolumn{1}{|c|}{139} 
& \multicolumn{1}{c|}{264} 
& \multicolumn{1}{c|}{800} 
& \multicolumn{1}{c|}{1000} 
& \multicolumn{1}{c|}{10} 
& \multicolumn{1}{c|}{800} \\\hline

\end{tabular}
\vskip0.2cm
\begin{tabular}{|c||c|c|c|} \hline
 $M^2_{D\alpha\beta}$
& \multicolumn{1}{c|}{\scriptsize{${\beta=1}$}} 
& \multicolumn{1}{c|}{\scriptsize{$\beta=2$}} 
& \multicolumn{1}{c|}{\scriptsize{$\beta=3$}} \\\hline\hline
 \scriptsize{$\alpha=1$}
& \multicolumn{1}{c|}{$(830)^2$} 
& \multicolumn{1}{c|}{0} 
& \multicolumn{1}{c|}{0} \\\hline

 \scriptsize{$\alpha=2$}
& \multicolumn{1}{c|}{0} 
& \multicolumn{1}{c|}{$(820)^2$} 
& \multicolumn{1}{c|}{0} \\\hline

 \scriptsize{$\alpha=3$}
& \multicolumn{1}{c|}{0} 
& \multicolumn{1}{c|}{0} 
& \multicolumn{1}{c|}{$(810)^2$} \\\hline
\end{tabular}
\hspace{0.6cm}
\begin{tabular}{|c||c|c|c|} \hline
 $M^2_{U\alpha\beta}$
& \multicolumn{1}{c|}{\scriptsize{${\beta=1}$}} 
& \multicolumn{1}{c|}{\scriptsize{$\beta=2$}} 
& \multicolumn{1}{c|}{\scriptsize{$\beta=3$}} \\\hline\hline
 \scriptsize{$\alpha=1$}
& \multicolumn{1}{c|}{$(820)^2$} 
& \multicolumn{1}{c|}{0} 
& \multicolumn{1}{c|}{0} \\\hline

 \scriptsize{$\alpha=2$}
& \multicolumn{1}{c|}{0} 
& \multicolumn{1}{c|}{$(600)^2$} 
& \multicolumn{1}{c|}{$(373)^2$} \\\hline

 \scriptsize{$\alpha=3$}
& \multicolumn{1}{c|}{0} 
& \multicolumn{1}{c|}{$(373)^2$} 
& \multicolumn{1}{c|}{$(580)^2$} \\\hline
\end{tabular}
\vskip0.4cm
\caption{\label{eberl:tab1}
Input parameters for the reference scenario
(mass parameters in GeV).
All $T_{U \a\b}$ and $T_{D \a\b}$ are zero.
}
\end{center}
\end{table}
In this scenario one has $\delta^{uLL}_{23}=0.068$, $\delta^{uRR}_{23}=0.4$ and 
$\delta^{uRL}_{23}=\delta^{uRL}_{32}=0$ for the QFV parameters. 
We obtain (masses in GeV):\\
$m_{\tilde u_1} = 472$, $m_{\tilde u_2} = 708$, 
$\tilde u_1 \sim -0.72\,\tilde c_R + 0.70\, \tilde t_R$, $\tilde u_2 \sim  0.70\, \tilde c_R + 0.71\, \tilde t_R$,
$m_{\tilde \chi_1^0} = 138$, and the branching ratios
$B(\ti{u}_1 \to c \nt_1)= 0.59, ~B(\ti{u}_1 \to t \nt_1)= 0.39, 
~B(\ti{u}_2 \to c \nt_1)= 0.44, ~B(\ti{u}_2 \to t \nt_1)= 0.40$. 
Note that the branching ratios of the decays of a squark into quarks of different 
generations are very large simultaneously, which could lead to large QFV effects. 
In our scenario this is a consequence of the fact that both squarks $\su_{1,2}$ are 
mainly strong mixtures of $\tilde c_R$ and $\tilde t_R$ due to the large 
$\tilde c_R -\tilde t_R$ mixing term $M^2_{U 23} (= (373 \gev)^2)$
and that $\nt_1$ is mainly the $U(1)$ gaugino. 
This also suppresses the couplings of $\su_{1,2}$ to $\tilde \chi^0_2$ and 
$\tilde \chi^+_1$ which are mainly $SU(2)$ gauginos. 
Note that $\nt_{3,4}$ and $\ch_2$ are very heavy in this scenario.

\begin{figure}
\begin{center}
\vspace{-3mm}
\scalebox{0.45}[0.73]{\includegraphics{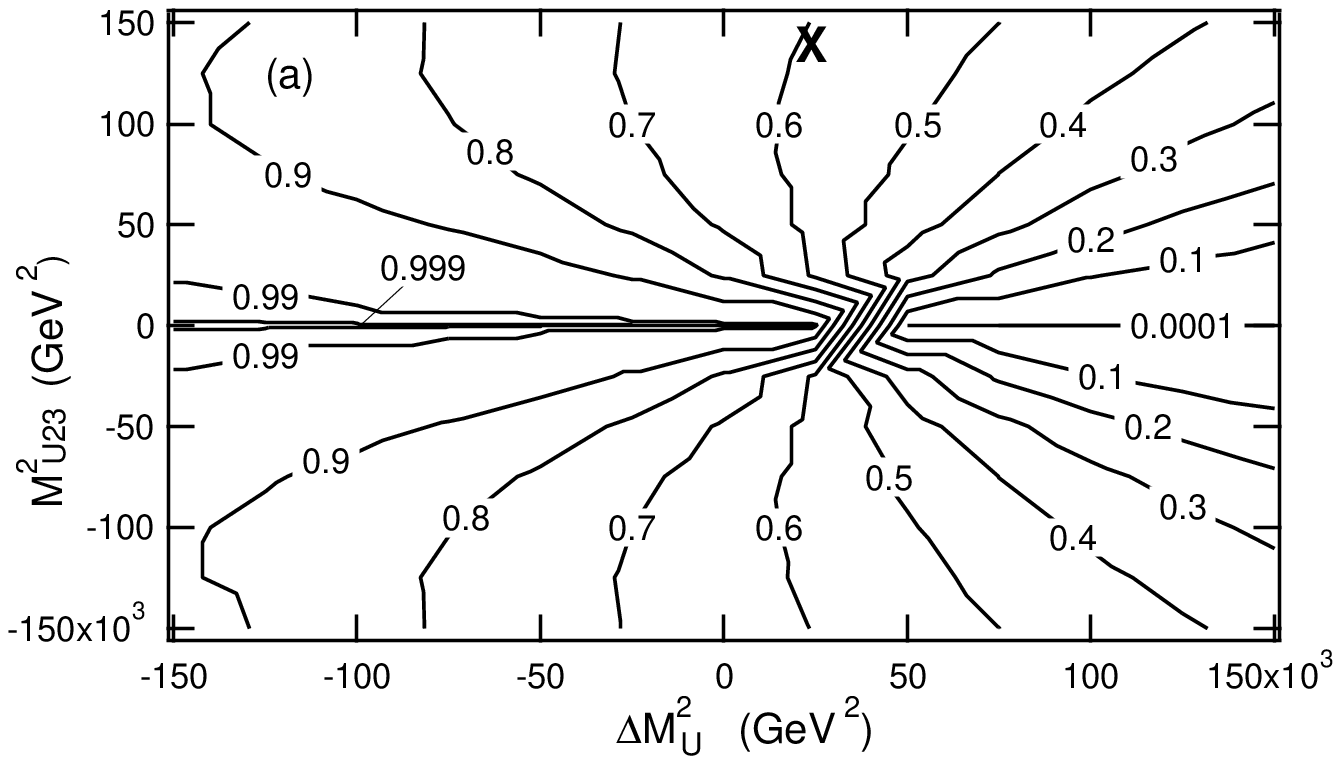}}  \hspace{5mm}
\scalebox{0.45}[0.73]{\includegraphics{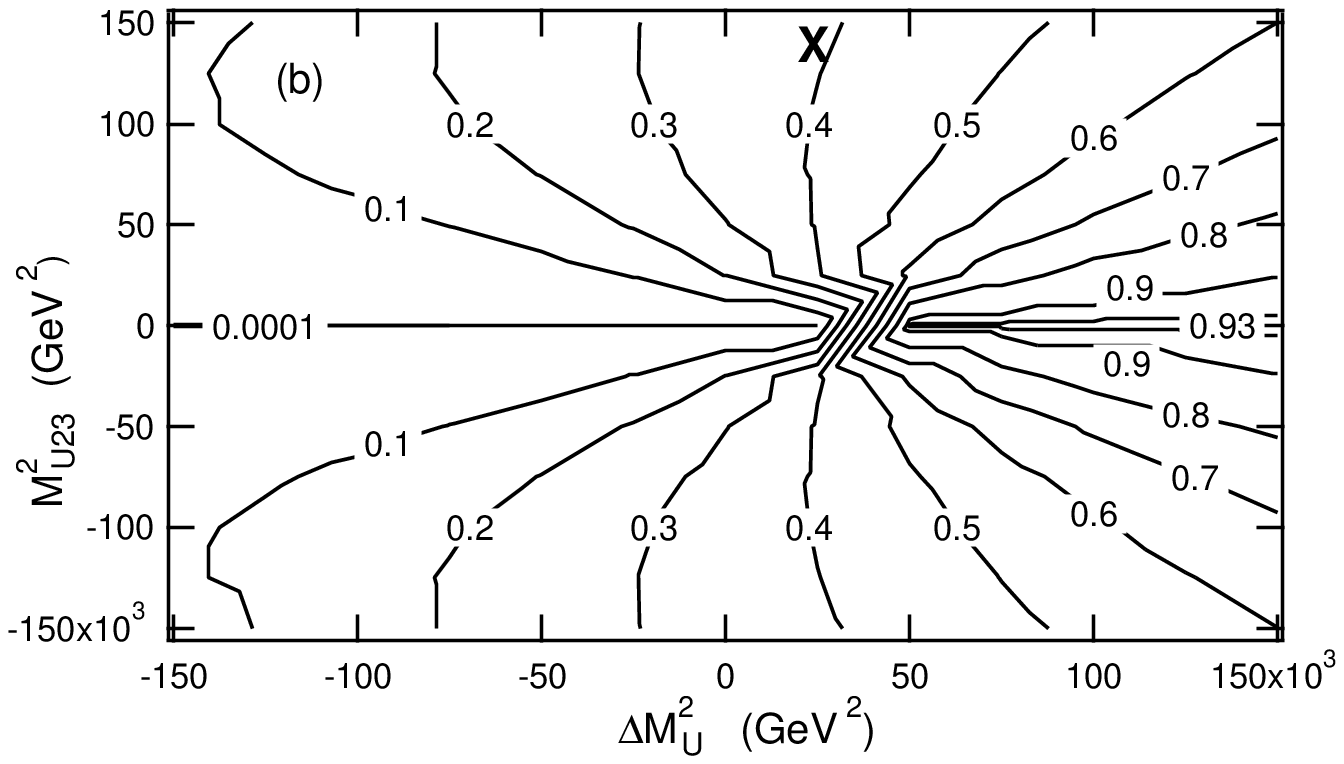}}
\caption{
Contours of QFV decay branching ratios (a)$B(\su_1 \to c \nt_1)$ and 
(b) $B(\su_1 \to t \nt_1)$ in the ($\Delta M^2_U, M^2_{U 23}$) plane.}
\label{eberl:fig1}
\vspace*{-2mm}
\end{center}
\end{figure}

Fig.~\ref{eberl:fig1} shows the contours of $B(\su_1 \to c \nt_1)$ and $B(\su_1 \to t \nt_1)$ 
in the ($\Delta M^2_U, M^2_{U 23}$) plane with $\Delta M^2_U \equiv 
M^2_{U 22} - M^2_{U 33}$. The range of $M^2_{U23}$ shown corresponds to the 
range $|\delta^{uRR}_{23}| < 0.45$ for $\Delta M^2_U=0$. 
In all plots the point corresponding to the scenario given in Table~\ref{eberl:tab1} is marked by an "x". 
We see that there are sizable regions where both decay modes are important 
at the same time. The observed behavior can be easily understood in the 
limit where the $\st_L$ - $\st_R$ mixing is neglected since in this limit only 
the mixing between $\sc_R$ and $\st_R$ is relevant for $\su_{1,2}$ and the 
corresponding effective mixing angle is given by  
$\tan(2\theta^{eff}_{\sc_R \st_R}) \equiv 2M^2_{U 23}/(\Delta M^2_U - m^2_t)$.
Note that for $\Delta M^2_U - m^2_t > 0$ [$\Delta M^2_U  - m^2_t < 0$], we have  
$\su_1$ $\sim$ $\st_R$ (+ $\sc_R$) [$\su_1$ $\sim$ $\sc_R$ (+ $\st_R$)].
We further study effects of the squark generation mixing on QFV signals at LHC. 
The large $B(\su_i \to c \nt_1)$ and $B(\su_i \to t \nt_1)$ $(i=1,2)$ may result 
in a sizable rate for the following QFV signals: 
\beq \label{eq:QFVsq_production_decay}
p~p \to \su_{1,2} ~\bar{\su}_{1,2} ~X \to c ~\bar{t} ~\nt_1 ~\nt_1 ~X, ~t ~\bar{c} ~\nt_1 ~\nt_1 ~X,
\eeq
where X contains only beam jets and the $\nt_1$'s give rise to missing transverse energy $E_T^{mis}$.
Defining
\beq
  \sigma^{ij}_{c t} \equiv 
  \sigma(pp \to \su_i \bar{\su}_j X) \times  \left(B(\su_i \to c \nt_1) \times B(\bar{\su}_j \to \bar{t} \nt_1) +
                                                                   B(\su_i \to t \nt_1) \times B(\bar{\su}_j \to \bar{c} \nt_1)  \right)\, ,
\eeq
we obtain the following cross sections 
at $E_{cm}$=14 TeV [7 TeV] for the scenario of Table \ref{eberl:tab1}: \\
$\sigma^{11}_{ct}$ = 172.8 [11.8] fb, $\sigma^{22}_{ct}$ = 11.5 [0.41] fb. 

\begin{figure}
\begin{center}
\vspace{-3mm}
\scalebox{0.5}[0.6]{\includegraphics{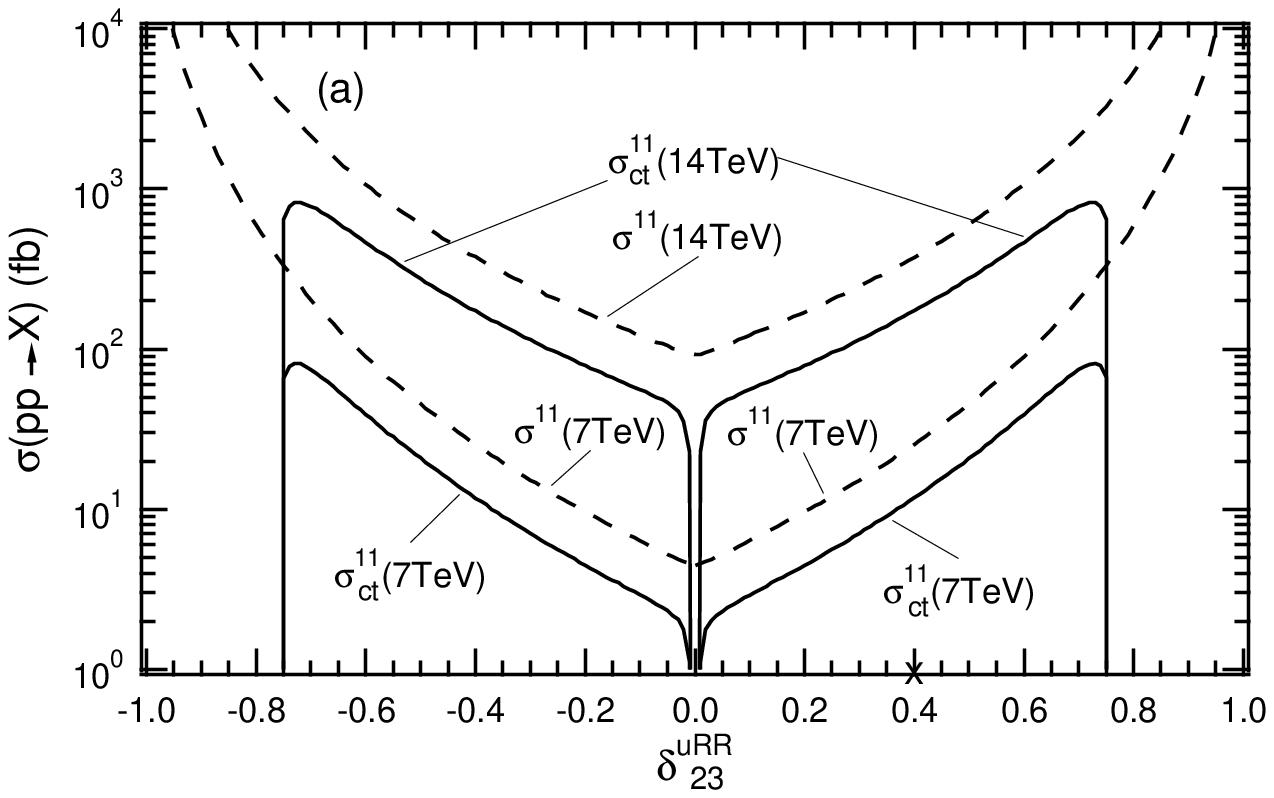}}
\scalebox{0.47}[0.6]{\includegraphics{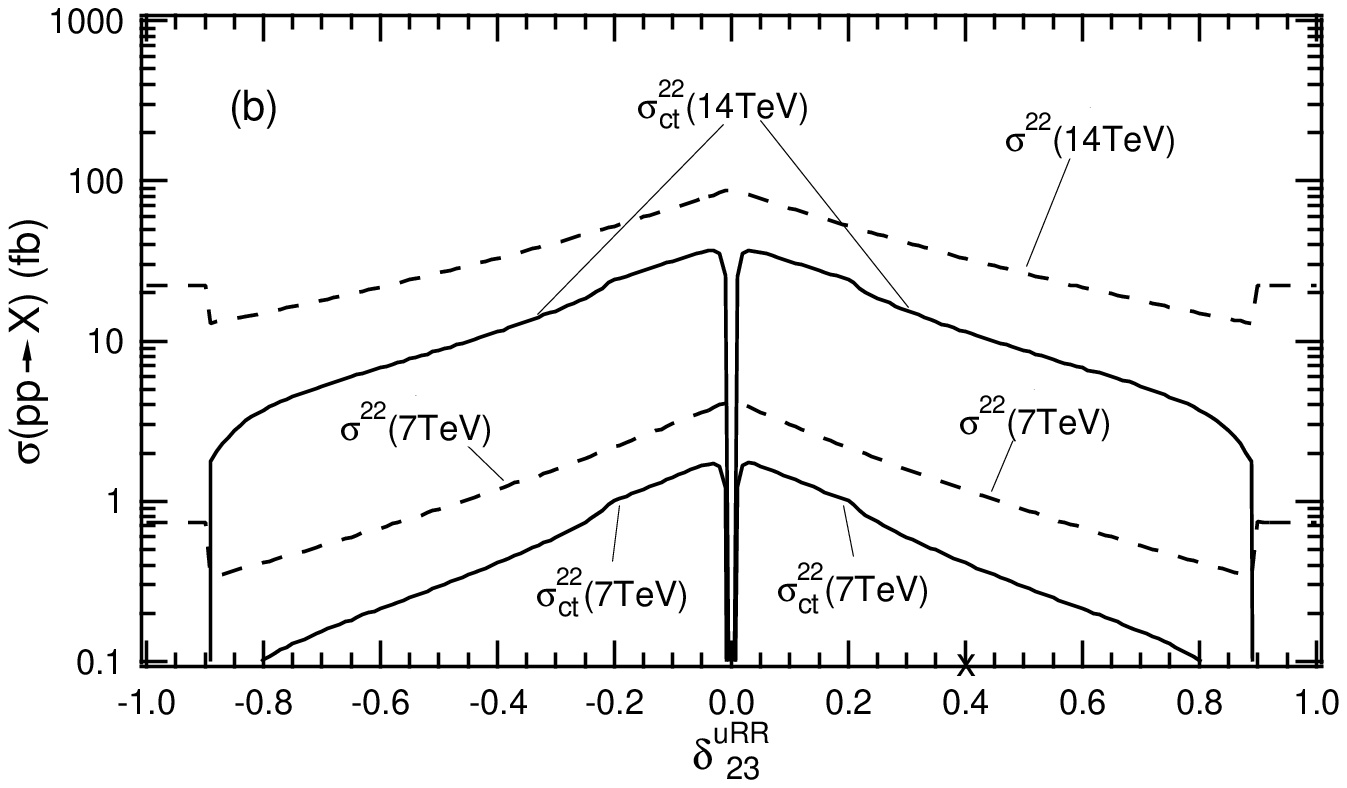}}
\caption{
$\delta^{uRR}_{23}$ dependences of (a) $\sigma^{11} \equiv \sigma(pp \to \su_1 \bar{\su}_1 X)$, 
$\sigma^{11}_{ct}$ and (b) $\sigma^{22} \equiv \sigma(pp \to \su_2 \bar{\su}_2 X)$, 
$\sigma^{22}_{ct}$.}
\label{eberl:fig2}
\vspace*{-2mm}
\end{center}
\end{figure}

In Fig.~\ref{eberl:fig2} we show the $\delta^{uRR}_{23}$ 
dependences of the QFV production cross sections $\sigma^{ii}_{ct}$ $(i=1,2)$ at $E_{cm}$ = 7 
and 14 TeV, where all basic parameters other than $M^2_{U 23}$ are fixed as in the scenario of 
Table \ref{eberl:tab1}. The QFV cross sections at 14 TeV are about an order of magnitude larger 
than those at 7 TeV. 
We see that the QFV cross sections quickly increase with increase of 
the QFV parameter $|\delta^{uRR}_{23}|$ around $\delta^{uRR}_{23} = 0$ and that they can be 
quite sizable in a wide allowed range of $\delta^{uRR}_{23}$. The mass of $\su_1$ ($\su_2$) 
decreases (increases) with increase of $|\delta^{uRR}_{23}|$. This leads to the increase of 
$\sigma^{11}_{ct}$ and the decrease of $\sigma^{22}_{ct}$ with increase of $|\delta^{uRR}_{23}|$.
$\sigma^{11}_{ct}$ vanishes for $|\delta^{uRR}_{23}| \gsim 0.76$, where the decay 
$\su_1 \to t \nt_1$ is kinematically forbidden. 
We have $\su_2 = \su_R$ for $|\delta^{uRR}_{23}| 
\gsim 0.9$, which explains the enhancement of $\sigma(pp \to \su_2 \bar{\su}_2 X)$  and the 
vanishing of $\sigma^{22}_{ct}$ for $|\delta^{uRR}_{23}| \gsim 0.9$. 
We see that the expected number of signal events at the LHC would be up to about
20000~(10) for an integrated luminosity of 100 fb$^{-1}$~(1 fb$^{-1}$) at 
$\sqrt{s}$ = 14 TeV (7 TeV).
Concerning the detectability, top-quark identification is necessary to distinguish the proposed
signal from $t \bar t$ production including missing energy. Efficient
charm-tagging would be very helpful for detecting flavour mixing between
the second and third generation. Otherwise, one should rather search for the signature ${\rm jet} + t\,(\bar t) + E_T^{\rm miss}$\cite{Bartl:QFV_squark}.\\[-3mm]

We remark that for the QFV scenarios based on a mSUGRA scenario such as SPS1a'
we have obtained results similar to the presented one\cite{Bartl:QFV_squark}.\\[-3mm]

In summary, despite the strong constraints from experimental data, non-minimal QFV can lead
to new signatures at LHC. This has been shown explicitly for one scenario. The given conclusions
hold, however, for a wide range of the MSSM parameter space\cite{Bartl:QFV_squark}.
Of course, the next necessary step is a detailed Monte-Carlo study including background reactions
and detector simulation
in order to identify the parameter region where the proposed QFV signal is observable 
with sufficient significance.  

\section{Acknowledgements}

The authors acknowledge support 
from the "Fonds zur F\"orderung der wissenschaftlichen Forschung" of Austria, project No.~P~18959-N16
and  project No.~I~297-N16, and from the DFG, project nr. PO-1337/2-1. 
This work is supported in part by the Landes-Exzellenzinitiative Hamburg.


\begin{thebibliography}{[1]}

\bibitem{Bartl:QFV_squark}%
A. Bartl, H. Eberl, B. Herrmann, K. Hidaka, W. Majerotto, W. Porod,
Phys. Lett. B 698 (2011) 380 [arXiv:1007.5483 [hep-ph]]; Erratum to be published.
   
\bibitem{Bartl:refstherein}%
See corresponding references in \cite{Bartl:QFV_squark}.

\end{thebibliography}
\end{document}